\documentclass[11pt]{article}
\usepackage[a4paper]{geometry}
\usepackage{natbib}
\usepackage{graphicx}
\usepackage{hyperref}
\usepackage{textcomp}
\usepackage{xcolor}
\usepackage{ulem}
\usepackage{lineno}
\usepackage[caption = false]{subfig}
\usepackage{amssymb}
\usepackage{amsmath}

\def\aap{Astron.\ \& Astrophys.}

\def\apjs{Astrophys.\ J.\ Suppl.}
\def\apjl{Astrophys.\ J. Letters}

\newcommand{\D}{\mathrm{d}}

\newcommand{\EQ}[1] {Eq.~(\ref{#1})}
\newcommand{\FIG}[1] {Figure~\ref{#1}}
\newcommand{\TAB}[1] {Table~\ref{#1}}

\newcommand{\AFF}[1]{$^{\foreach\d[count=\ni]in{#1}{\ifnum\ni=1\ref{\d}\else,\ref{\d}\fi}}$}

\def\cmpc{\mathrm{pc}\,\mathrm{cm}^{-3}}

\definecolor{dkblue}{RGB}{54, 86, 169}

\newcommand{\footremember}[2]{%
    \footnote{#2}
    \newcounter{#1}
    \setcounter{#1}{\value{footnote}}%
}
\newcommand{\footrecall}[1]{%
    \footnotemark[\value{#1}]%
} 

\begin{document}

\title{Periodic variation of magnetoionic environment of a fast radio burst source}

\author{Jiangwei Xu\footremember{naoc}{National Astronomical Observatories, Chinese Academy of Sciences, 20A Datun Road, Chaoyang District, Beijing 100101, People's Republic of China}
\and
Heng Xu \footrecall{naoc}
\and
Yanjun Guo \footrecall{naoc}
\footremember{mpifr}{Max-Planck institut für Radioastronomie, Auf Dem H\"ugel, Bonn, 53121, Germany}
\and Jinchen Jiang \footrecall{mpifr}
\and Bojun Wang\footrecall{naoc}
\author{Zihan Xue}
\footremember{pku}{Department of Astronomy, Peking University, Beijing 100871, People's Republic of China}
\footremember{kiaa}{Kavli Institute for Astronomy and Astrophysics, Peking University, Beijing 100871, People's Republic of China}
\and
Yunpeng Men\footrecall{mpifr}
\and
Kejia Lee\footrecall{pku}\footrecall{naoc}\footremember{ynao}{Yunnan Observatories, Chinese Academy of Sciences, Kunming 650216, People's Republic of China}\footremember{bla}{Beijing Laser Acceleration Innovation Center, Huairou, Beijing, 101400, People's Republic of China}
\and
Bing Zhang\footremember{ULNV}{The Nevada Center for Astrophysics, University of Nevada, Las Vegas, NV 89154, USA}
\footremember{ULNVA}{Department of Physics and Astronomy, University of Nevada, Las Vegas, NV 89154, USA}
\and
Weiwei Zhu\footrecall{naoc}
\and
Jinlin Han \footrecall{naoc}
}
\maketitle

\begin{abstract}
Fast radio bursts (FRBs) are luminous, dispersed millisecond-duration radio bursts whose origin is poorly known\citep{Petroff19AAR,Cordes19ARAA,Zhang20Nature}. 
Recent observations\citep{CHIME20Nat,Xu2022Nature,Mckinven2023ApJ,Anna2023Science,Li2025FAST} suggest that some FRBs may reside in binary systems\citep{2020ApJ...893L..26I,Wang2022NatCo}, even though conclusive evidence remains elusive.
Here we report the detection of a 26.24$\pm$0.02 day periodicity in Faraday rotation measure (RM) of an actively repeating source named FRB~20201124A. The detection was made from 3,106 bursts collected with the Five-hundred-meter Aperture Spherical radio Telescope (FAST) over $\sim$ 365 days. The RM periodicity is coherently phase-connected across $\sim$14 cycles over a 1-year duration. Our detection of RM periodicity corresponds to a natural logarithmic Bayesian factor of 1,168. The detection significances vary between 5.9$-$34 $\sigma$ under different assumptions.
Such a periodicity provides evidence for the binary nature of FRB~20201124A, where the periodic RM variations arise from the orbital motion of the FRB source within the magnetoionic environment of the system. 
Together with previous observations\citep{CHIME20Nat,Xu2022Nature,Mckinven2023ApJ,Anna2023Science,Li2025FAST}, our result suggests that being in binary systems may be a common feature for actively repeating FRB sources. 

\end{abstract}



\section{Maintext} 
The repeating FRB~20201124A\citep{Xu2022Nature,Jiang2022RAA,CHIMEFRB20201124A,Kirsten2024NatAs} resides in a compact persistent radio source region\citep{2024Natur.632.1014B} in the galaxy SDSS~J050803.48+260338.0 at a distance\citep{Xu2022Nature} of $\sim450\,\mathrm{Mpc}$. The radio bursts of FRB~20201124A are believed to be of a magnetospheric origin as indicated by the observed polarization angle swings\citep{Luo2020Nature, 2025Natur.637...48N,Xu2022Nature,2024ApJ...972L..20N} and  90\% circular polarization of one burst\citep{Jiang2024NSR}. As the linearly polarized emission of FRB travels through the environment, magnetic fields and free electrons induce the Faraday rotation effect, namely, a frequency-dependent rotation of linear polarization plane defined by the position angle (PA) in the form of $\Delta {\rm PA}={\rm RM}\, \lambda^2$, with $\lambda$ being the electromagnetic wave wavelength.
The value of RM depends on the magnetic field and electron density in the environment with $\mathrm{RM}=\int_0^d n_e(l) B_{||}(l) dl$, where the integration is performed along the line-of-sight (LOS), $B_{||}(l)$ is the parallel magnetic field strength and $n_e(l)$ is the electron density. The RM variation of the source indicates that the source is enclosed by an au-scale magnetoionic environment\citep{Xu2022Nature}.
For FRB~20201124A, the dynamical evolution of the magnetoionic environment was revealed by the highly varied RM (by $\sim 500 {\rm cm^{-3}\, pc}$ over a month) as well as the frequency-dependent oscillation structures in burst polarization intensity\citep{Xu2022Nature}.

The intense activity of FRB~20201124A provides a unique opportunity to study the mechanism of prominent variations of RM in FRBs. Three distinct activity cycles were identified in FRB~20201124A following its discovery\citep{CHIMEFRB20201124A,Xu2022Nature,Jiang2022RAA,Kirsten2024NatAs}. The first cycles (from April to May, 2021) ended with a rapid cessation of burst activity\citep{Xu2022Nature,Zhang2022RAA}. Significant variation of RM was observed during the first activity cycle\citep{Xu2022Nature}. The brevity of the second activity cycle lasted for four days in September 2021, during which the RM displayed a decreasing trend\citep{Zhang2022RAA}. During the last activity cycle (from February to April, 2022), the FRB source also showed RM variations similar to the first cycle, albeit with a markedly reduced amplitude. To better understand the variability and quantify any underlying periodicities, we conducted several periodic analyses on the RM data.

\begin{figure} 
\centering
\includegraphics[width=\textwidth]{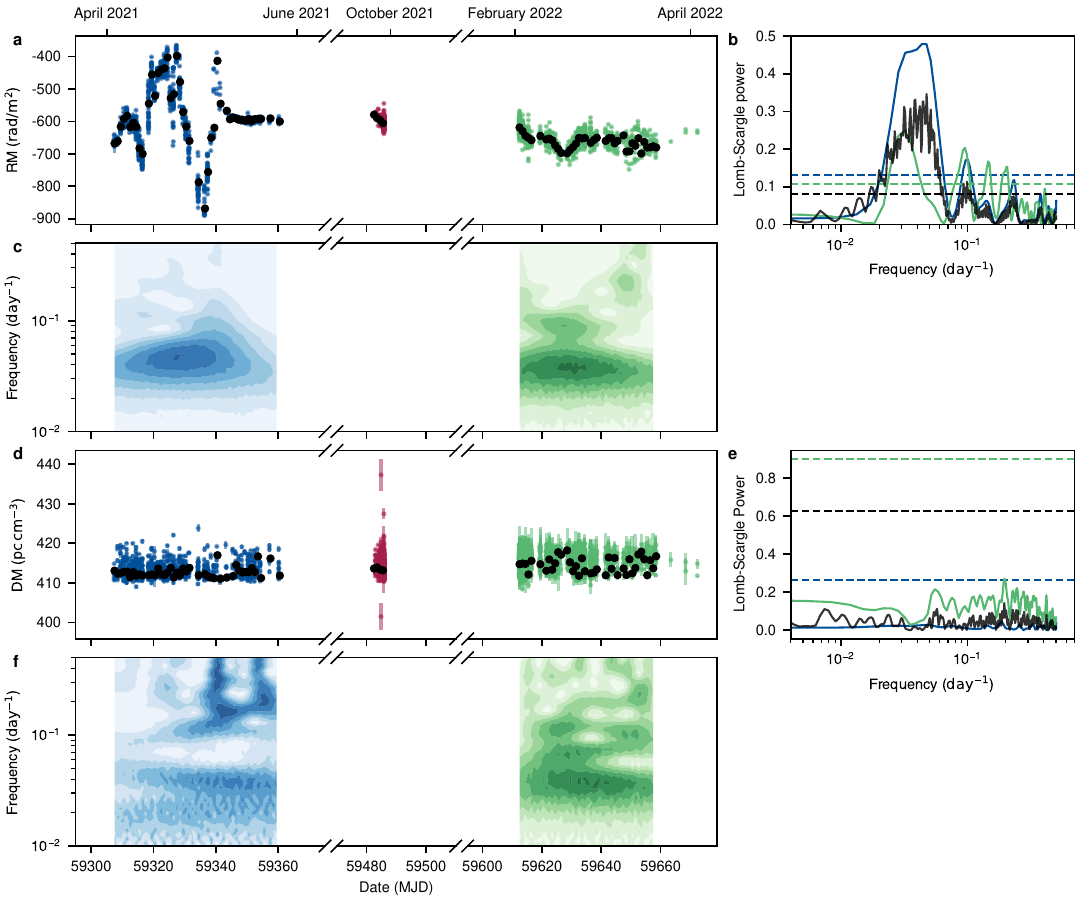}
\caption{\textbf{Periodicity analysis of RM and DM variations of FRB~202011244A.} {\bf a}, RM measurements as a function of time, where blue, red and green data are from refs. \citep{Xu2022Nature,Jiang2022RAA} and this work, and the black dots represents the daily averaged data. {\bf b}, Lomb-Scargle periodogram of the RM measurements, with the first epoch, third epoch, and all epochs shown as blue, green and black curves, respectively, the horizontal dashed lines indicate the false-alarm probability of $10^{-6}$. {\bf c}, Wavelet analysis of the daily averaged RM values. {\bf d}, {\bf e}, and {\bf f}, The same as {\bf (a)}, {\bf (b)}, and {\bf (c)},  but for the DM measurement, respectively.
\label{fig:fig1}
}
\end{figure}

We searched for periodicity using the Lomb-Scargle periodogram, which is effective in identifying sinusoidal signal with unknown frequency in the unevenly sampled data. The details of computing the Lomb-Scargle periodogram are in the Methods. 
The results for the first and third cycles, as well as the entire RM dataset, are in \FIG{fig:fig1}(panel {\bf b}). Prominent peaks were identified at periods of $\sim28$ and $\sim10$ days for the first epoch, and $\sim34$ and $\sim11$ days for the third epoch. The uncertainties represent the full width at half maximum of the highest value of each peak. In the periodogram for the entire RM dataset, RM variation show significant power in the range of $\sim20-30$ days (with multiple narrow peaks whose significances are larger than $33\sigma$), along with power around $\sim10$ days. These features suggest the presence of an underlying periodicity that may be persistent across the full 1-yr timespan.

We then performed wavelet analyses on the daily average RM data during the first and third activity cycles to explore the stationarity of the variation period (see Methods) as shown in \FIG{fig:fig1} (panel {\bf c}). The wavelet power for the two epochs reveal stationary frequencies of $\sim0.04\,\mathrm{day^{-1}}$ (corresponding to $\sim25\,\mathrm{days}$) spanning the entire duration of the 1st and 3rd active cycles. The frequencies of the two cycles remain stable over time and are consistent with each other. The wavelet analyses confirm the periodic signals in the RM data as revealed by the Lomb-Scargle analyses. Furthermore, the wavelet analyses indicates that the periodicity is stationary in time rather than limited to individual time windows. 

\begin{figure}
\centering
\includegraphics[width=1.\textwidth]{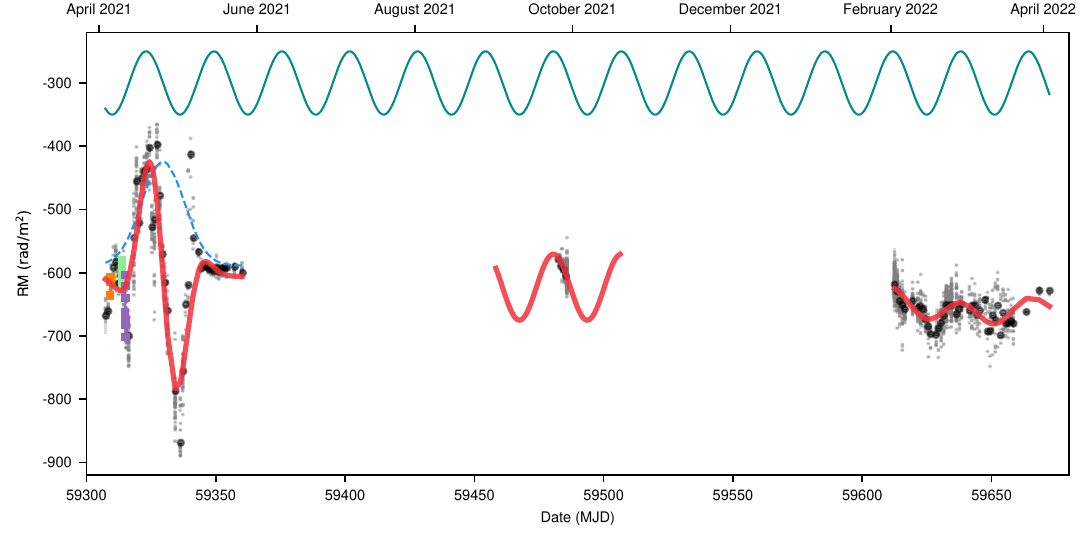}
\caption{\textbf{Fitting of the temporal variations of RM of FRB~20201124A.} The RM measurements are shown in grey dots, with the daily averaged values shown in black. The best-fitting result using \EQ{eq:RM_curve} is present as red curve. The Gaussian envelop in the first epoch is plotted in blue dashed curve. A sinusoidal curve across the whole time span using the best-fit values of period and reference phase is plotted as green solid curve in arbitrary unit, and it was shifted to show that the RM variation is coherently phase-connected. The RM data from Effelsberg\citep{2021MNRAS.508.5354H}, Parkes\citep{2022MNRAS.512.3400K} and GBT\citep{2022Sci...375.1266F} are shown in light green, orange and purple respectively, which agree with the trend in FAST data.
\label{fig:waveform}
}
\end{figure}

To study the phase coherency of RM variations, we fit a single sinusoidal waveform with amplitude modulation to the RM of all three cycles.
We found that the RM variation can be well described by the following model
\begin{equation}
{\rm RM}(t) = 
    \begin{cases}
    A_1\cos{(2\pi f t +\phi)}\, e^{\frac{-|t-t_1|^2}{\tau_1^2}}+ b_1 ,& \textrm{for the 1st cycle}\,\, 59300<t<59370 \\
    A_2\cos{(2\pi f t +\phi)} + b_2 ,& \textrm{for the 2nd cycle}\,\,59482<t<59486 \\
    A_3\cos{(2\pi f t +\phi)} + b_3+c_3t+d_3t^2 ,& \textrm{for the 3rd cycle}\,\,59600<t<59680
    \end{cases}
    \label{eq:RM_curve}
\end{equation}
where the sinusoidal variation in the first activity cycle is modulated by a Gaussian function with amplitude $A_1$, width $\tau_1$, and centered at $t_1$. The RM variation of the third activity cycle is modeled with sinusoidal waveform with a constant amplitude ($A_3$) and quadratic baseline (i.e. parameter $b_3$, $c_3$,and $d_3$). $f$ and $\phi$ are the frequency and initial phase of the sinusoid, respectively. In the second activity cycle, no modulation is needed due to the short timespan. The Bayesian inference is applied to find the global best fitted parameters as presented in \FIG{fig:waveform}. This periodic model fits the data well over 14 turns. The details of the Bayesian inference is in Methods. 
The best-fit frequency is $0.03811 \pm 0.00003 \,{\rm day}^{-1}$, corresponding to a period of $26.24\pm0.02$ days. The Bayesian evidence of the periodic model is 1,168 compared to the white-noise only model, and the likelihood ratio test gives a significance of 34$\sigma$. 

A more conservative estimation for the significance of the periodicity comes from the inclusion of the potential systematics. We compare the change in the maximum likelihood values $\Delta \ln\mathcal{L}$ when fitting RM variation using the sinusoidal signal in \EQ{eq:RM_curve} with different frequencies, and calculate the significance of the best-fit frequency according to the null distribution of $\Delta \ln\mathcal{L}$ (see Methods for details). This gives a p-value of $2.0\times10^{-9}$, or a significance level of $5.9\sigma$.

The fitting residuals of the above single-frequency model show high-frequency signatures after subtracting the best-fit curve from the RM measurements. The Lomb-Scargle analysis and wavelet analysis also indicate more than one frequency components. We found that by introducing another sinusoidal signal component with a higher frequency, one can further phase-coherently fit the data of all three epochs. The model with two frequency components (see Methods) produced the best-fitting of the second frequency being $0.08915\pm0.00004\,{\rm day}^{-1}$, or a period of $11.217\pm0.005\,{\rm day}$.
The fitting residuals of the dual-frequency model still show structures in the first active cycle, which is mainly due to the over-simplification of the Gaussian amplitude modulation assumption.

Given that the observed RM variation could originate from either LOS magnetic field or electron density fluctuations, we further investigated the temporal behavior of DM. No significant periodicity are found in DM variations using either the Lomb-Scargle periodogram nor wavelet analysis (panels {\bf e}, and {\bf f} in \FIG{fig:fig1}). A very loose constraint could be set on the upper limit of DM variation that $\delta {\rm DM}< 5\, \mathrm{cm^{-3}\,pc}$. We note that a DM variation of $\sim0.1\, \mathrm{cm^{-3} pc}$ with a milli-Gauss-level magnetic field is sufficient to generate the RM variations with the observed amplitudes (i.e. ${\rm RM}\simeq810\, {\rm DM/pc\cdot cm^{-3}} B/{\rm mG}$). However, for a micro-Gauss level interstellar magnetic field, the required DM variation is $\sim 100\,\mathrm{cm^{-3}\,pc}$, much larger than the current DM variation constraint.

The observed RM periodicity is inconsistent with scenarios involving intervening magnetized environments, ionized screens within the FRB host galaxy or the Milky Way\citep{1996ApJ...458..194M,2023MNRAS.520.2039Y}. Such scenarios typically produce stochastic RM variations or long-term monotonic trends rather than coherent periodic signals in the RM. The periodic RM variation is inconsistent with the prediction of an isolated supernova remnant, which results in random RM variations in the short term and a secular RM decrease in the long term\citep{Piro2018,2023MNRAS.520.2039Y}.

The detection of a 26.24-day periodicity in the rotation measure (RM) variations of the repeating FRB 20201124A provides critical insights into the nature of its progenitor system. This periodicity echoes the 16-day activity cycle observed in FRB 20180916B\citep{CHIME20Nat}, which has been interpreted as evidence for either binary orbital motion or precession of an isolated neutron star. However, the RM periodicity in FRB 20201124A favors the binary origin\citep{2020ApJ...893L..26I}, as the precession scenario is inconsistent with the observed RM variations.
In the precession model, geometric changes in the magnetosphere would not induce significant RM variability. While precession alters the magnetospheric orientation relative to Earth, the line-of-sight direction remains fixed, leaving the RM contribution from the surrounding circumsource environment unaffected. Furthermore, RM contributions intrinsic to the magnetosphere are negligible: relativistic motion of electrons/positrons and the presence of pair plasma suppress measurable Faraday rotation\citep{2011MNRAS.417.1183W,2023MNRAS.520.2039Y}. Thus, the observed RM periodicity likely reflects changes in the magnetized environment external to the neutron star, such as those driven by orbital motion in a binary system.

The observed RM periodicity is most plausibly explained by a binary system comprising a compact star (e.g., a neutron star or a black hole) and a companion that carries a magnetized ionized environment. 
Within this framework, the FRB emission originates from the compact star traversing through the local magnetoionic medium, which imparts RM modulation by the orbital motion of the compact star with respect to the companion. The binary system also explains the RM amplitude modulation (the Gaussian) on the periodic variation (the sinusoidal). The sinusoidal variation is induced from binary motion, changing the magnetic field geometric configuration 
with respect to the line of sight; and the Gaussian modulation comes from the plasma density fluctuations caused by the companion's activities, such as stellar wind or jet ejection.
The companion star can be a main sequence star\citep{Wang2022NatCo}, a white dwarf \citep{2016ApJ...823L..28G}, a neutron star \citep{2020ApJ...890L..24Z}, or even a massive black hole\citep{2018ApJ...854L..21Z}.

In the past, RM variations had been observed in low-mass companion systems such as spider binary pulsars\citep{2020MNRAS.495.3052C,2023Natur.618..484L,2023ApJ...955...36W}. 
FRB~20201124A is unlikely in a spider pulsar system, of which the orbital period would be too short. 
The Kepler's law suggests $M=330 M_{\odot} (a/{\rm au})^3 (p/{\rm 20 \, day})^{-2}$, where $M$ is the binary system total mass and $a$ is the semi-major separation. Thus, a stellar mass binary system is more plausible assuming a sub-au $a$\citep{2023MNRAS.520.2039Y,2023A&A...673A.136R}, where the RM variation is introduced by the plasma casting from a massive companion star, such as circumstellar disk\citep{Wang2022NatCo} or wind/outflow\citep{2023MNRAS.520.2039Y}.

The periodicity in FRB 20201124A can be further verified through future detections of bursts with polarization observations. While this RM periodicity remains unique among FRBs observed to date, RM variations are common in repeating sources\citep{Mckinven2023ApJ, Anna2023Science, Li2025FAST}, suggesting that similar periodicities could emerge in other FRBs as observational datasets grow.
However, such RM variations may not manifest universally. For example, they could be undetectable in face-on binary systems (where orbital motion induces minimal line-of-sight environmental changes) or in systems at evolutionary stages distinct from FRB~20201124A, where a cleaner intra-binary environment (e.g., reduced magnetized material) suppresses measurable RM fluctuations. Regardless, systematic, high-cadence polarization observations of FRBs will remain a critical diagnostic for probing progenitor systems and their magneto-ionic environments.

\section{Supplementary materials} 

\subsection{Observation and data}\label{sec:data}

We continued monitoring FRB~20201124A using the FAST radio telescope\citep{Jiang19SCPMA} after its reactivation in 2022\citep{2022ATel_Atri}. We observed it from 2nd February 2022 to 26 May 2022 with a total observing time of 34.3 hours. The detailed observation setups and data reduction procedures basically follow our previous works\citep{Xu2022Nature,Zhou2022RAA,Zhang2022RAA,Jiang2022RAA,Niu2022RAA}, and will be described in a parallel paper. Below, we provide a concise summary of the observations and data processing relavent to RM and DM measurements, readers are directed to those previous studies for methodological specifics\citep{Xu2022Nature,Zhou2022RAA,Zhang2022RAA,Jiang2022RAA,Niu2022RAA}.

The search mode data recorded using a time resolution of 49.152 $\mathrm{\mu s}$ and 4,096 channels covering the frequency range of 1-1.5 GHz. Burst candidates with threshold of S/N>7 were found using single pulse searching software \textsc{transientx}\citep{2024A&A...683A.183M} and then visually identified, bursts data were then polarization calibrated. The RM values were estimated for 1,467 bursts with S/N$>$20, using a Q-U fitting method\citep{DKL19}, which can provide a robust estimation on the uncertainties. The DMs were derived by maximising the time derivative of intensity in the Fourier domain\citep{2018Natur.553..182M} to align structures in dynamic spectra of FRBs due to their complex morphology, rather than traditional maximum-S/N techniques used in the pulsar studies. The DM values for 1,270 bursts were obtained.

We incorporated the RM and DM measurements from our previous studies, i.e. 1,103 RM and 811 DM measurements from the first activity cyle, 536 RM from the second activity cycle\citep{Xu2022Nature,Jiang2022RAA}. We calculated the DM of the bursts in ref. \citep{Jiang2022RAA} using the same method, and 514 DM measurements were obtained. In total, this analysis utilized 3,106 RM and 2,595 DM measurements.

\subsection{Lomb-Scargle periodogram}\label{sec:lsp_ana}
We performed periodicity analyses on the RM dataset using the Lomb-Scargle periodogram (LSP) algorithm\citep{Lomb1976,Scargle1982}. The LSPs were generated as follows: 1) a second-degree polynomial baseline was fitted to model long-term trend and subtracted from the dataset, 2) the normalized periodogram was computed across a frequency range of 0.004 to 0.5 day$^{-1}$. To aid the inspection of LSP results, we calculated the threshold power at a false-alarm probability (FAP) of $10^{-6}$ through the bootstrap method. We applied the lombscargle tool in the open-source package \textsc{astropy} to compute normalized periodogram and threshold power. We first analyzed the RM data from the first and third activity cycles independently. The second cycle was excluded due to insufficient temporal coverage (spanning only a few days). The resulting periodograms are shown in \FIG{fig:fig1}b. In the first activity cycle, two prominent broad peaks above the threshold power were identified at $\sim$0.04 and $\sim$0.095 $\mathrm{day^{-1}}$, with FAPs of $10^{-151}$ and $2\times10^{-41}$, translating to significances of $26.2\sigma$ and $13.4\sigma$. The FAPs were estimated using a conservative method proposed by ref.\citep{2008MNRAS.385.1279B} since the FAPs are too small to use bootstrap method. The third activity cycle exhibited four peaks above the threshold power at $\sim$0.03, $\sim$0.095, $\sim$0.145, and $\sim$0.2 $\mathrm{day^{-1}}$, with FAPs of $7\times10^{-87}$, $2\times10^{-68}$, $3\times10^{-54}$, and $7\times10^{-49}$ (equivalent to $19.8\sigma$, $17.3\sigma$, $15.4\sigma$, and $14.7\sigma$). Notably, the two lowest-frequency peaks demonstrate consistency between the first and third cycles, implying a underlying common signal. We then compute the LSP for the whole RM dataset, which is also shown is \FIG{fig:fig1}b. The LSP revealed multiple broad bumps, where each contains multiple narrow peaks. The two leftmost bumps align with the low-frequency peaks observed in the first and third cycles. The peaks on the left bump show FAPs below $4\times10^{-239}$ ($33\sigma$), with the three highest peaks occurring at frequencies of 0.0386, 0.0446, and 0.0474 $\mathrm{day^{-1}}$, having FAPs of $2\times10^{-266}$, $10^{-261}$ and $2\times10^{-281}$ (corresponding to significances of $34.8\sigma$, $34.5\sigma$, and $35.8\sigma$). Importantly, the 0.0386 $\mathrm{day^{-1}}$ signal matches the results of our model curve fitting (see details in Section of RM fitting).

The same analysis was applied to the DM dataset (\FIG{fig:fig1}e). No statistically significant periodic signals were detected, as none of the peaks in the periodograms of the first and third activity cycles or the whole DM dataset exceeded the threshold power of FAPs of $10^{-6}$. Burst-to-burst DM variations are evident in the data, which we attribute to a combination of 1) complex burst morphology, 2) limitations in revolving high-time-resolution fine structures for burst alignment, since only incoherent dedispersion can be applied currently. We derive a conservative estimate of $\Delta \mathrm{DM}\lesssim 10\,\cmpc$ for the DM variations on the timescale of $\sim10$ days.

\subsection{Wavelet analysis}\label{sec:wavelet_ana}
In order to analyze distribution of the power of the underlying periodicities,  we performed periodicity analysis using the wavelet analysis\citep{Torrence1998} on the RM and DM datasets. The wavelet power spectra were generated as follows: 1) the measurements were daily averaged to match the requirement that the wavelet analysis needs evenly sampled data, we interpolated the daily averaged data since the FAST observation were not ideally evenly spaced. 2) the 2-D wavelet power spectra was calculated by applying the continuous wavelet transform using a complex Morlet wavelet. We employed the python software package \textsc{pywt} to conduct the continuous wavelet transform. Like the way we did in Lomb-Scargle analyses, we produced the wavelet power spectra of RM and DM measurements for the first and third activity cycles independently, the second cycle was not analyzed for the same reason aforementioned. The daily averaged RM and DM data on days with less than 3 measurements were discarded in this analysis, to minimize potential biases arising from insufficient sample sizes, since RM and DM measurements in our data set showing large burst-to-burst variations. For the same reason, the averaged data of the last three days with RM measurements and the averaged data of the last two days with DM measurements are excluded. The results are shown in \FIG{fig:fig1}c and \FIG{fig:fig1}f. The wavelet power spectra of RM datasets of the first and the third cycles both display significant $\sim0.04\,{\rm day^{-1}}$ signals across the entire time spans, and no significant signals can be seen in the wavelet power spectra of DM datasets. The frequencies of signals in the RM dataset are consistent with the LSP result. The wavelet analysis results indicate that those signals are persistent over time during the first and third active cycles.

\subsection{RM fitting}\label{sec:rmfit}

For quantitative characterization of the RM variation, we perform curve fitting using the formula in \EQ{eq:RM_curve} to fit the variation of RM in the three epochs coherently. 
There are complex short-term variations of RM in excess of the fluctuations described by measurement noise. We add Equad parameters to the formal uncertainty of RM measurements to account for these short-term variations. Equad parameter is widely used in the noise analyses of pulsar timing studies to model white noise in pulsar timing residuals\citep{Lentati2014} by adding additional uncertainty to the data error in quadrature.
The likelihood function (the probability distribution of the data given a set of parameters) is
\begin{equation}
    \mathcal{L} = \frac{1}{\prod_i{\sqrt{2\pi (\sigma_i^2+\rm{Equad(t_i)}^2)}}} \exp{\left[-\frac{1}{2}\frac{({\rm RM}_i-\rm{sig}(t_i))^2}{\sigma_i^2+\rm{Equad(t_i)}^2}\right]}\,
\end{equation}
where $RM_i$, $\sigma_i$, and $t_i$, are the RM value, RM measurement uncertainty and time of arrival in the Solar barycentric frame of the $i$th burst in our dataset, $\mathrm{sig}(t)$ is the signal function defined by \EQ{eq:RM_curve}, and ${\rm Equad}(t)$ takes different values ${\rm Eq}_1$,${\rm Eq}_2$ and ${\rm Eq}_3$ for three epochs respectively. 

\begin{figure}
\centering
\includegraphics[width=\textwidth]{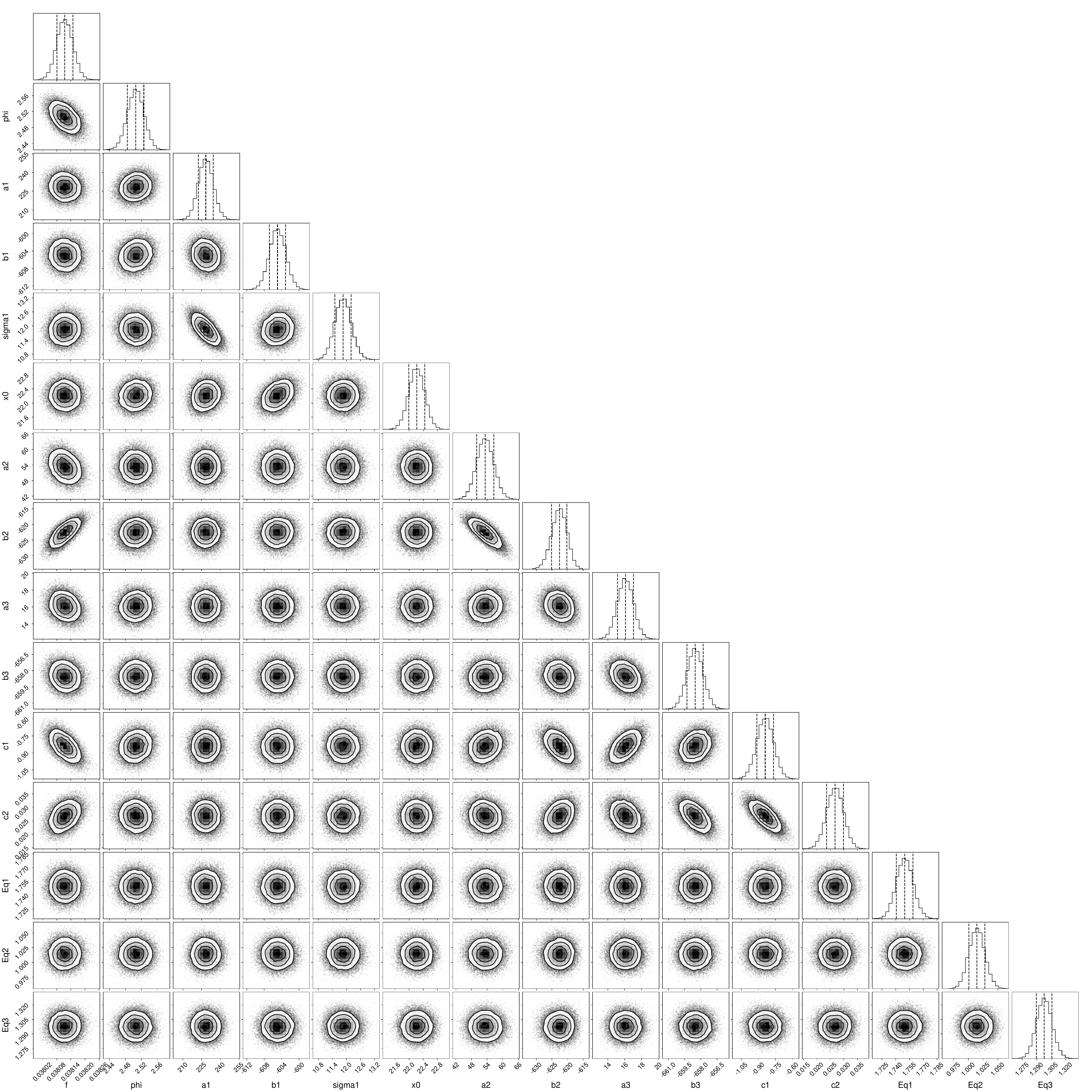}
\caption{\textbf{The posterior distribution of the parameters for modulated sinusoidal signal in \EQ{eq:RM_curve}.} The plots on the diagonal are marginalized distribution for each parameter, while off-diagonal plots shows the covariances between two parameters.
\label{fig:posterior}
}
\end{figure}

We perform Bayesian analysis to do parameter estimation and model comparison. Given data ${\boldsymbol d}$, the probability distribution of the parameters $p({\boldsymbol \theta}|{\boldsymbol d})$ is related to the likelihood via the Bayes theorem,
\begin{equation}
    p({\boldsymbol \theta}|{\boldsymbol d})=\frac{p({\boldsymbol d}|{\boldsymbol \theta})p({\boldsymbol \theta})}{p({\boldsymbol d})},
\end{equation}
where $p({\boldsymbol d}|{\boldsymbol \theta})=\mathcal{L}$ is the likelihood function, $p({\boldsymbol \theta})$ is the prior distribution of the parameters, and $p({\boldsymbol d})$ is the marginal likelihood, or Bayes evidence.
Bayes evidence acts as a normalization factor for parameter estimation, but is central to model selection.
The Bayes Factor $\mathcal{B}$ is defined as the ratio between the Bayes evidence of two models. If $\mathcal{B}_{M_1}^{M_2}>1$, the model $M_2$ is preferred over $M_1$ given the data.
We employed nested sampling method which could give estimation of Bayesian evidence, and the sampler used is \textsc{MultiNest}\citep{2009MNRAS.398.1601F}. The prior distribution of the parameters in \EQ{eq:RM_curve} and Equad are summarized in \TAB{tab:prior}.
The posterior distribution of these fitting parameters are shown in \FIG{fig:posterior}. The best-fit frequency is $0.03811 \pm 0.0003 \,{\rm day}^{-1}$, corresponding to a period of $26.24\pm0.02$ days. The reconstructed waveform using the best-fit parameters is present in \FIG{fig:waveform}.

\begin{table}
    \centering
    \begin{tabular}{c|c|c}
    \hline
    \hline
        Parameter & Prior distribution & Prior range \\
    \hline
        $f$ & Uniform & $[0.025,0.055]$   \\
        $\phi$ & Uniform & $[0,2\pi]$ \\
        $A_1$ & Uniform & $[0,1000]$ \\
        $\tau_1$ & Uniform & $[0,100]$ \\
        $b_1$ & Uniform & $[-1000,1000]$ \\
        $t_1$ & Uniform & $[0,50]$ \\
        $A_2$ & Uniform & $[0,1000]$ \\
        $b_2$ & Uniform & $[-1000,1000]$ \\
        $A_3$ & Uniform & $[0,1000]$ \\
        $b_3$ & Uniform & $[-1000,1000]$ \\
        $c_3$ & Uniform & $[-10,10]$ \\
        $d_3$ & Uniform & $[-10,10]$ \\
        ${\rm Equad}$ & Log10-Uniform & $[-2,3]$ \\
    \hline
    \hline
    \end{tabular}
    \caption{Prior distribution of the parameters in RM fitting.}
    \label{tab:prior}
\end{table}

\begin{figure}
\centering
\includegraphics[width=\textwidth]{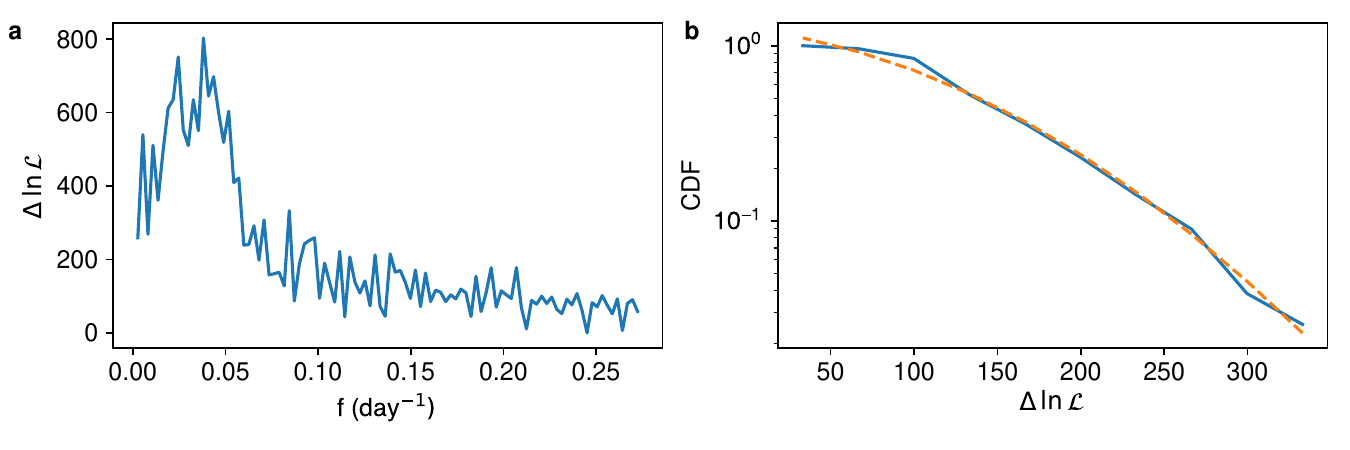}
\caption{\textbf{Significance evaluation based on the likelihood ratio at different frequencies.} \textbf{a)} : the likelihood ratio $\Delta \ln \mathcal{L}$ when fitting RM variation with a modulated sin signal in \EQ{eq:RM_curve} fixing at different frequencies. \textbf{b)}: cumulative distribution of $\Delta \ln \mathcal{L}$ (blue solid line) and an analytical function fitting the distribution (orange dashed line).
\label{fig:chi2}
}
\end{figure}

\begin{figure}
\centering
\includegraphics[width=\textwidth]{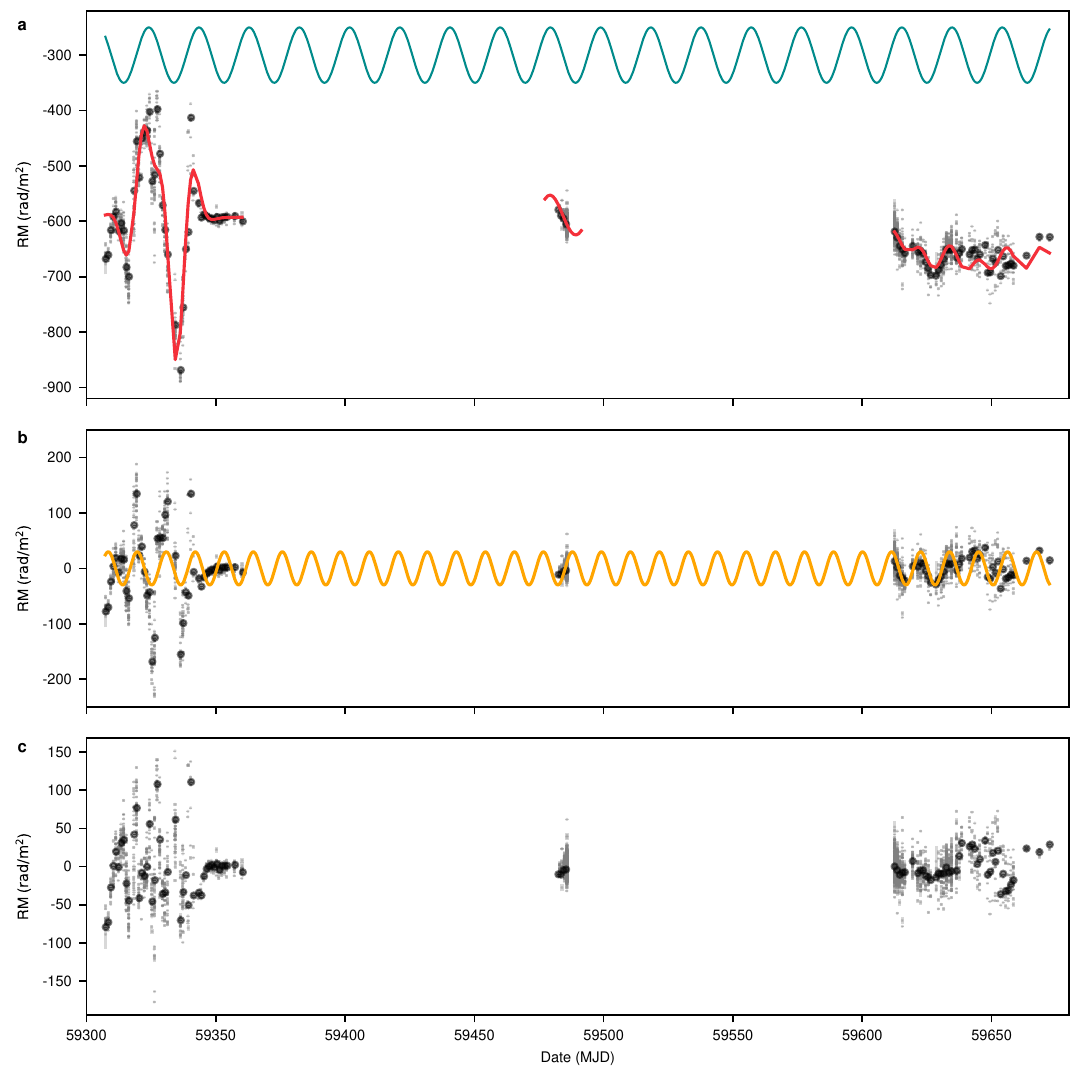}
\caption{\textbf{Fitting of the temporal RM variations of FRB~20201124A using a two-components modulated sinusoidal signal}. \textbf{a}, The original RM variation shown in black dots and the fitted waveform using two-frequency model shown in red. The low-frequency sinusoidal signal (in arbitrary unit and shifted) is plotted in green for reference. \textbf{b}, The residuals after subtracting the low-frequency part of the model are in black dots, and the high-frequency sinusoidal signal is plotted in orange to show the phase coherency with the residuals. \textbf{c}: The RM residuals after subtracting the best-fit waveform of the full two-frequency model.
\label{fig:waveform_2sin}
}
\end{figure}

To estimate the significance of the periodic signal, we take the model with white noise only as null hypothesis. This white noise model include Equad parameters and constant $b$ for each epoch to fit the baseline, and the likelihood function is 
\begin{equation}
    \mathcal{L}_{\rm w} = \frac{1}{\prod_i{\sqrt{2\pi (\sigma_i^2+\rm{Equad}^2)}}} \exp{\left[-\frac{1}{2}\frac{({\rm RM}_i-b)^2}{\sigma_i^2+\rm{Equad}^2}\right]}\,.
\end{equation}  
We compare the Bayesian evidence for the model with modulated periodic signal and the null hypothesis. The natural logarithmic Bayesian factor ($\ln \mathcal{B}$) is 1,168, which indicates a strong preference for the proposed model with modulated periodic signal.
We also calculate the likelihood ratio $\Delta \ln\mathcal{L}=1216$, which is the ratio between the maximum likelihood values of the two models. The likelihood ratio follows $\chi^2$ distribution with N degree of freedom, where $N=9$ is the difference between the parameter dimensions of the two models. The FAP is $4\times10^{-256}$, corresponding to a significance of $34\sigma$. We also check the significance of the periodicity in the first and third epochs respectively. For the first epoch, $\Delta \ln\mathcal{L}=707$, and the significance is $26 \sigma$ (FAP of $2\times10^{-150}$); for the third epoch, $\Delta \ln\mathcal{L}=392$, and the significance is $19 \sigma$ (FAP of $2\times10^{-82}$).

Considering the complex structure of RM variation, there could be significant systematic effects. Take white noise scenario as null hypothesis would overestimate the significance when there are additional noise components.
We estimate the significance of the detected period using another conservative method which takes account of the systematic effects of the data.
We compare the change in the maximum likelihood $\mathcal{L}_{\rm MLE}$ when fitting the data using signal with different frequencies, and use the values of $\Delta \ln\mathcal{L}$ at the high frequency tail to estimate the null distribution. The values of $\Delta \ln\mathcal{L}$ for different frequencies are shown in \FIG{fig:chi2}a. We use the $\Delta \ln\mathcal{L}$ for frequencies larger than 0.06\,day$^{-1}$ to approximate the null distribution, as shown in \FIG{fig:chi2}b. Extrapolating the distribution to the $\Delta \ln\mathcal{L}$ value at the peak frequency results in a p-value of $2.0\times10^{-9}$, and the significance is $5.9\sigma$.

The amplitude-modulated sinusoidal signal captures the dominant RM variability features, and we establish statistical significance of the detected periodicity using this single-frequency model. Subtraction of the best-fit waveform revealed residual power at higher frequencies. To account for these systematic deviations, we implemented a dual-frequency model incorporating an additional sinusoidal component. We perform similar former discussed Bayesian analysis using this two-frequency model, and the fitting result is shown in \FIG{fig:waveform_2sin}. The best-fit second frequency is $0.08915\pm0.00004\,{\rm day}^{-1}$, and the corresponding period is $11.217\pm0.005\,{\rm day}$.
The phase evolution of the secondary component shows consistency with the residual structure across the full data span (Fig. \FIG{fig:waveform_2sin}b). Residuals from the dual-frequency fit (Fig. \FIG{fig:waveform_2sin}c) exhibit reduced scatter, but retain coherent structures likely attributable to the complex amplitude modulation.

\subsection{Burst activity dependence on periodic phase}

\begin{figure}
\centering
\includegraphics[width=\textwidth]{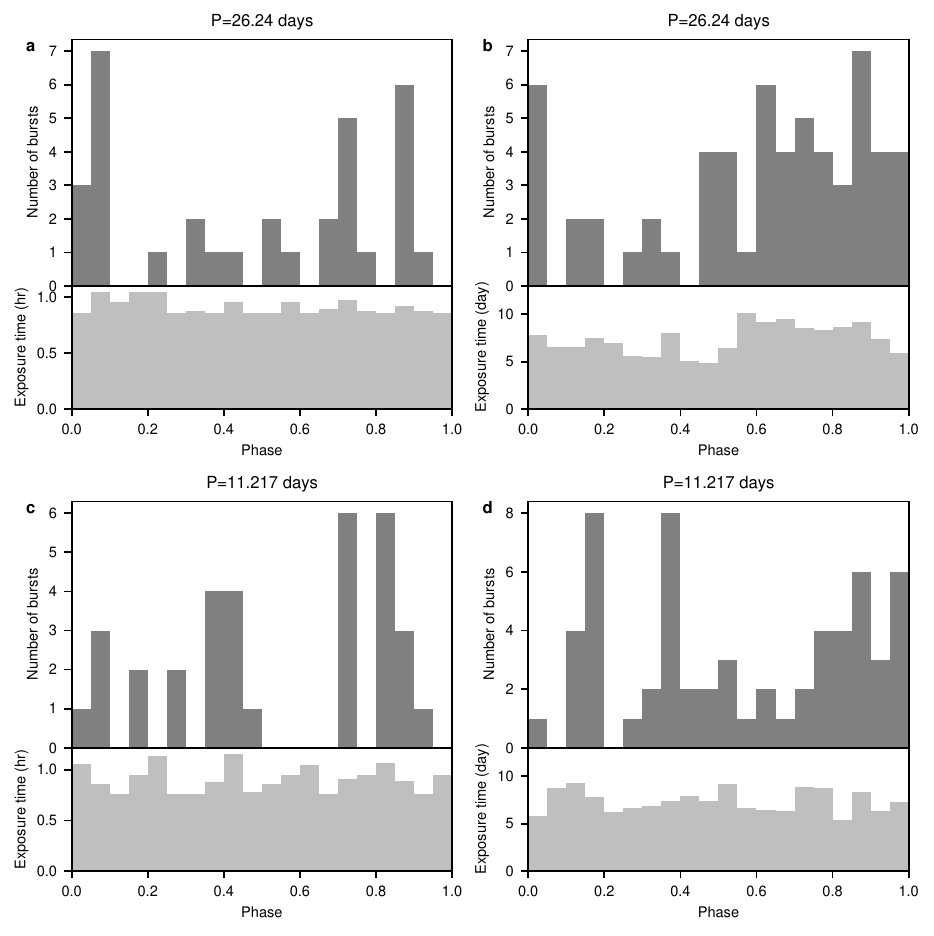}
\caption{\textbf{Burst activities against phase.} \textbf{a}, CHIME sample folded with period of 26.24-day. \textbf{c}, CHIME sample folded with period of 11.217-day. \textbf{b}, burst sample from ref. \citep{Kirsten2024NatAs} folded with period of 26.24-day. \textbf{d}, burst sample from ref. \citep{Kirsten2024NatAs} folded with period of 11.217-day. In each figure, upper panel shows the number of bursts at each phase, and lower panel show the exposure time at each phase.
\label{fig:burstratephase_other}
}
\end{figure}

We investigated the phase dependence of burst activity for the two identified periods (26.24-day and 11.217-day). We phase-folded the bursts from two datasets: 1) 33 bursts detected by CHIME\citep{CHIMEFRB20201124A}, covering the first activity cycle, and 2) 46 bursts detected by four small dishes described in ref. \citep{Kirsten2024NatAs}, spanning all three activity cycles. Both datasets are well-suited for this analysis due to their extended monitoring baseline and large burst counts. The folded results with 20 phase bins are shown in \FIG{fig:burstratephase_other}. For 20 phase bins, the maximal derived phase uncertainties induced by frequency measurement errors are approximately approximately 0.01 and 0.01 for the 26.24-day and 11.217-day periods respectively, which can be negligible for this analysis. For the CHIME sample, the reduced $\chi^2$ values are 2.6 and 2.5 for the 26.24-day and 11.217-day periods, respectively. For the data in ref.\citep{Kirsten2024NatAs}, the reduced $\chi^2$ are 1.5 and 1.9 for the two periods. Therefore we conclude that no evident periodic burst activities can be found with the two periods, both for very bright and faint bursts.

\section{Acknowledgments}
We thank L. Spitler for constructive discussions.
This work made use of the data from FAST (Five-hundred-meter Aperture Spherical radio Telescope, \url{https://cstr.cn/31116.02.FAST}) operated by National Astronomical Observatories, Chinese Academy of Sciences.
This work is supported by the National SKA Program of China (2020SKA0120100), Natural Science Foundation of China (12403093,11988101), and the Chinese Academy of Sciences via project JZHKYPT-2021-06. KJL acknowledges supports from the XPLORER PRIZE. 

\section{Contributions}

H.X., Y.J.G., K.J.L., and B.Z co-supervised data analyses and interpretations, and led the paper writing. J.W.X., and J.C.J. conducted the RM and DM measurements. H.X., Y.J.G, B.J.W., Z.H.X., and K.J.L. performed the timing analysis and periodicity search. Y.P.M. contributed to the searching software development. J.L.H. and W.W.Z. coordinated the observation.



%
\subsubsection*{Data availability}
The directly related data that support the findings of this study can be found at the PSRPKU website, \url{https://psr.pku.edu.cn/index.php/publications/frb20201124a/}.

\subsubsection*{Code availability}
\noindent
\textsc{pywt} (\url{https://github.com/PyWavelets/pywt})

\noindent
\textsc{astropy} (\url{https://www.astropy.org})

\noindent
\textsc{pymultinest} (\url{https://github.com/JohannesBuchner/PyMultiNest})





\bibliography{ms}{}

\begin{thebibliography}{}
\expandafter\ifx\csname natexlab\endcsname\relax\def\natexlab#1{#1}\fi
\providecommand{\url}[1]{\href{#1}{#1}}
\providecommand{\dodoi}[1]{doi:~\href{http://doi.org/#1}{\nolinkurl{#1}}}
\providecommand{\doeprint}[1]{\href{http://ascl.net/#1}{\nolinkurl{http://ascl.net/#1}}}
\providecommand{\doarXiv}[1]{\href{https://arxiv.org/abs/#1}{\nolinkurl{https://arxiv.org/abs/#1}}}

\bibitem[{{Anna-Thomas} {et~al.}(2023){Anna-Thomas}, {Connor}, {Dai}, {Feng},
  {Burke-Spolaor}, {Beniamini}, {Yang}, {Zhang}, {Aggarwal}, {Law}, {Li},
  {Niu}, {Chatterjee}, {Cruces}, {Duan}, {Filipovic}, {Hobbs}, {Lynch}, {Miao},
  {Niu}, {Ocker}, {Tsai}, {Wang}, {Xue}, {Yao}, {Yu}, {Zhang}, {Zhang}, {Zhu},
  \& {Zhu}}]{Anna2023Science}
{Anna-Thomas}, R., {Connor}, L., {Dai}, S., {et~al.} 2023, Science, 380, 599,
  \dodoi{10.1126/science.abo6526}

\bibitem[{{Atri} {et~al.}(2022){Atri}, {Bilous}, {van Leeuwen}, {Maan},
  {Pastor-Marazuela}, {Petroff}, \& {Straal}}]{2022ATel_Atri}
{Atri}, P., {Bilous}, A., {van Leeuwen}, J., {et~al.} 2022, The Astronomer's
  Telegram, 15197, 1

\bibitem[{{Baluev}(2008)}]{2008MNRAS.385.1279B}
{Baluev}, R.~V. 2008, \mnras, 385, 1279,
  \dodoi{10.1111/j.1365-2966.2008.12689.x}

\bibitem[{{Bruni} {et~al.}(2024){Bruni}, {Piro}, {Yang}, {Quai}, {Zhang},
  {Palazzi}, {Nicastro}, {Feruglio}, {Tripodi}, {O'Connor}, {Gardini},
  {Savaglio}, {Rossi}, {Nicuesa Guelbenzu}, \&
  {Paladino}}]{2024Natur.632.1014B}
{Bruni}, G., {Piro}, L., {Yang}, Y.-P., {et~al.} 2024, \nat, 632, 1014,
  \dodoi{10.1038/s41586-024-07782-6}

\bibitem[{{Chime/Frb Collaboration} {et~al.}(2020){Chime/Frb Collaboration},
  {Amiri}, {Andersen}, {Bandura}, {Bhardwaj}, {Boyle}, {Brar}, {Chawla},
  {Chen}, {Cliche}, {Cubranic}, {Deng}, {Denman}, {Dobbs}, {Dong}, {Fandino},
  {Fonseca}, {Gaensler}, {Giri}, {Good}, {Halpern}, {Hessels}, {Hill},
  {H{\"o}fer}, {Josephy}, {Kania}, {Karuppusamy}, {Kaspi}, {Keimpema},
  {Kirsten}, {Landecker}, {Lang}, {Leung}, {Li}, {Lin}, {Marcote}, {Masui},
  {McKinven}, {Mena-Parra}, {Merryfield}, {Michilli}, {Milutinovic},
  {Mirhosseini}, {Naidu}, {Newburgh}, {Ng}, {Nimmo}, {Paragi}, {Patel}, {Pen},
  {Pinsonneault-Marotte}, {Pleunis}, {Rafiei-Ravandi}, {Rahman}, {Ransom},
  {Renard}, {Sanghavi}, {Scholz}, {Shaw}, {Shin}, {Siegel}, {Singh}, {Smegal},
  {Smith}, {Stairs}, {Tendulkar}, {Tretyakov}, {Vanderlinde}, {Wang}, {Wang},
  {Wulf}, {Yadav}, \& {Zwaniga}}]{CHIME20Nat}
{Chime/Frb Collaboration}, {Amiri}, M., {Andersen}, B.~C., {et~al.} 2020, \nat,
  582, 351, \dodoi{10.1038/s41586-020-2398-2}

\bibitem[{{Cordes} \& {Chatterjee}(2019)}]{Cordes19ARAA}
{Cordes}, J.~M., \& {Chatterjee}, S. 2019, ARA\&A, 57, 417,
  \dodoi{10.1146/annurev-astro-091918-104501}

\bibitem[{{Crowter} {et~al.}(2020){Crowter}, {Stairs}, {McPhee}, {Archibald},
  {Boyles}, {Hessels}, {Karako-Argaman}, {Lorimer}, {Lynch}, {McLaughlin},
  {Ransom}, {Roberts}, {Stovall}, \& {van Leeuwen}}]{2020MNRAS.495.3052C}
{Crowter}, K., {Stairs}, I.~H., {McPhee}, C.~A., {et~al.} 2020, \mnras, 495,
  3052, \dodoi{10.1093/mnras/staa933}

\bibitem[{{Desvignes} {et~al.}(2019){Desvignes}, {Kramer}, {Lee}, {van
  Leeuwen}, {Stairs}, {Jessner}, {Cognard}, {Kasian}, {Lyne}, \&
  {Stappers}}]{DKL19}
{Desvignes}, G., {Kramer}, M., {Lee}, K., {et~al.} 2019, Science, 365, 1013,
  \dodoi{10.1126/science.aav7272}

\bibitem[{{Feng} {et~al.}(2022){Feng}, {Li}, {Yang}, {Zhang}, {Zhu}, {Zhang},
  {Lu}, {Wang}, {Dai}, {Lynch}, {Yao}, {Jiang}, {Niu}, {Zhou}, {Xu}, {Miao},
  {Niu}, {Meng}, {Qian}, {Tsai}, {Wang}, {Xue}, {Yue}, {Yuan}, {Zhang}, \&
  {Zhang}}]{2022Sci...375.1266F}
{Feng}, Y., {Li}, D., {Yang}, Y.-P., {et~al.} 2022, Science, 375, 1266,
  \dodoi{10.1126/science.abl7759}

\bibitem[{{Feroz} {et~al.}(2009){Feroz}, {Hobson}, \&
  {Bridges}}]{2009MNRAS.398.1601F}
{Feroz}, F., {Hobson}, M.~P., \& {Bridges}, M. 2009, \mnras, 398, 1601,
  \dodoi{10.1111/j.1365-2966.2009.14548.x}

\bibitem[{{Gu} {et~al.}(2016){Gu}, {Dong}, {Liu}, {Ma}, \&
  {Wang}}]{2016ApJ...823L..28G}
{Gu}, W.-M., {Dong}, Y.-Z., {Liu}, T., {Ma}, R., \& {Wang}, J. 2016, \apjl,
  823, L28, \dodoi{10.3847/2041-8205/823/2/L28}

\bibitem[{{Hilmarsson} {et~al.}(2021){Hilmarsson}, {Spitler}, {Main}, \&
  {Li}}]{2021MNRAS.508.5354H}
{Hilmarsson}, G.~H., {Spitler}, L.~G., {Main}, R.~A., \& {Li}, D.~Z. 2021,
  \mnras, 508, 5354, \dodoi{10.1093/mnras/stab2936}

\bibitem[{{Ioka} \& {Zhang}(2020)}]{2020ApJ...893L..26I}
{Ioka}, K., \& {Zhang}, B. 2020, \apjl, 893, L26,
  \dodoi{10.3847/2041-8213/ab83fb}

\bibitem[{{Jiang} {et~al.}(2022){Jiang}, {Wang}, {Xu}, {Xu}, {Zhang}, {Wang},
  {Zhou}, {Zhang}, {Niu}, {Lee}, {Zhang}, {Han}, {Li}, {Zhu}, {Dai}, {Feng},
  {Jing}, {Li}, {Luo}, {Miao}, {Niu}, {Tsai}, {Wang}, {Wang}, {Xu}, {Yang},
  {Yang}, {Yao}, \& {Yuan}}]{Jiang2022RAA}
{Jiang}, J.-C., {Wang}, W.-Y., {Xu}, H., {et~al.} 2022, Research in Astronomy
  and Astrophysics, 22, 124003, \dodoi{10.1088/1674-4527/ac98f6}

\bibitem[{{Jiang} {et~al.}(2024){Jiang}, {Xu}, {Niu}, {Lee}, {Zhu}, {Zhang},
  {Qu}, {Xu}, {Zhou}, {Cao}, {Wang}, {Wang}, {Cao}, {Zhang}, {Zhang}, {Gan},
  {Han}, {Hao}, {Huang}, {Jiang}, {Li}, {Li}, {Li}, {Li}, {Luo}, {Men}, {Qian},
  {Sun}, {Wang}, {Xu}, {Xu}, {Yang}, {Yao}, {Yue}, {Yu}, {Yuan}, \&
  {Zhu}}]{Jiang2024NSR}
{Jiang}, J.~C., {Xu}, J.~W., {Niu}, J.~R., {et~al.} 2024, National Science
  Review, 12, nwae293, \dodoi{10.1093/nsr/nwae293}

\bibitem[{{Jiang} {et~al.}(2019){Jiang}, {Yue}, {Gan}, {Yao}, {Li}, {Pan},
  {Sun}, {Yu}, {Liu}, {Tang}, {Qian}, {Lu}, {Yan}, {Peng}, {Zhang}, {Wang},
  {Li}, \& {Li}}]{Jiang19SCPMA}
{Jiang}, P., {Yue}, Y., {Gan}, H., {et~al.} 2019, \scpma, 62, 959502,
  \dodoi{10.1007/s11433-018-9376-1}

\bibitem[{{Kirsten} {et~al.}(2024){Kirsten}, {Ould-Boukattine}, {Herrmann},
  {Gawro{\'n}ski}, {Hessels}, {Lu}, {Snelders}, {Chawla}, {Yang}, {Blaauw},
  {Nimmo}, {Puchalska}, {Wolak}, \& {van Ruiten}}]{Kirsten2024NatAs}
{Kirsten}, F., {Ould-Boukattine}, O.~S., {Herrmann}, W., {et~al.} 2024, Nature
  Astronomy, 8, 337, \dodoi{10.1038/s41550-023-02153-z}

\bibitem[{{Kumar} {et~al.}(2022){Kumar}, {Shannon}, {Lower}, {Bhandari},
  {Deller}, {Flynn}, \& {Keane}}]{2022MNRAS.512.3400K}
{Kumar}, P., {Shannon}, R.~M., {Lower}, M.~E., {et~al.} 2022, \mnras, 512,
  3400, \dodoi{10.1093/mnras/stac683}

\bibitem[{{Lanman} {et~al.}(2022){Lanman}, {Andersen}, {Chawla}, {Josephy},
  {Noble}, {Kaspi}, {Bandura}, {Bhardwaj}, {Boyle}, {Brar}, {Breitman},
  {Cassanelli}, {Dong}, {Fonseca}, {Gaensler}, {Good}, {Kaczmarek}, {Leung},
  {Masui}, {Meyers}, {Ng}, {Patel}, {Pearlman}, {Petroff}, {Pleunis},
  {Rafiei-Ravandi}, {Rahman}, {Sanghavi}, {Scholz}, {Shin}, {Stairs},
  {Tendulkar}, \& {Zwaniga}}]{CHIMEFRB20201124A}
{Lanman}, A.~E., {Andersen}, B.~C., {Chawla}, P., {et~al.} 2022, \apj, 927, 59,
  \dodoi{10.3847/1538-4357/ac4bc7}

\bibitem[{{Lentati} {et~al.}(2014){Lentati}, {Alexander}, {Hobson}, {Feroz},
  {van Haasteren}, {Lee}, \& {Shannon}}]{Lentati2014}
{Lentati}, L., {Alexander}, P., {Hobson}, M.~P., {et~al.} 2014, \mnras, 437,
  3004, \dodoi{10.1093/mnras/stt2122}

\bibitem[{{Li} {et~al.}(2023){Li}, {Bilous}, {Ransom}, {Main}, \&
  {Yang}}]{2023Natur.618..484L}
{Li}, D., {Bilous}, A., {Ransom}, S., {Main}, R., \& {Yang}, Y.-P. 2023, \nat,
  618, 484, \dodoi{10.1038/s41586-023-05983-z}

\bibitem[{{Li} {et~al.}(2025){Li}, {Zhang}, {Yang}, {Tsai}, {Yang}, {Law},
  {Anna-Thomas}, {Chen}, {Lee}, {Tang}, {Xiao}, {Xu}, {Yang}, {Chen}, {Feng},
  {Li}, {Mckinven}, {Niu}, {Shin}, {Wang}, {Zhang}, {Zhang}, {Zhou}, {Zhu},
  {Dai}, {Chang}, {Geng}, {Han}, {Hu}, {Li}, {Luo}, {Niu}, {Shi}, {Sun}, {Wu},
  {Zhu}, {Jiang}, \& {Zhang}}]{Li2025FAST}
{Li}, Y., {Zhang}, S.~B., {Yang}, Y.~P., {et~al.} 2025, arXiv e-prints,
  arXiv:2503.04727, \dodoi{10.48550/arXiv.2503.04727}

\bibitem[{{Lomb}(1976)}]{Lomb1976}
{Lomb}, N.~R. 1976, \apss, 39, 447, \dodoi{10.1007/BF00648343}

\bibitem[{{Luo} {et~al.}(2020){Luo}, {Wang}, {Men}, {Zhang}, {Jiang}, {Xu},
  {Wang}, {Lee}, {Han}, {Zhang}, {Caballero}, {Chen}, {Chen}, {Gan}, {Guo},
  {Hao}, {Huang}, {Jiang}, {Li}, {Li}, {Li}, {Luo}, {Pan}, {Pei}, {Qian},
  {Sun}, {Wang}, {Wang}, {Wen}, {Xu}, {Xu}, {Yan}, {Yan}, {Yu}, {Yuan},
  {Zhang}, \& {Zhu}}]{Luo2020Nature}
{Luo}, R., {Wang}, B.~J., {Men}, Y.~P., {et~al.} 2020, \nat, 586, 693,
  \dodoi{10.1038/s41586-020-2827-2}

\bibitem[{{Mckinven} {et~al.}(2023){Mckinven}, {Gaensler}, {Michilli}, {Masui},
  {Kaspi}, {Bhardwaj}, {Cassanelli}, {Chawla}, {Dong}, {Fonseca}, {Leung},
  {Li}, {Ng}, {Patel}, {Petroff}, {Pearlman}, {Pleunis}, {Rafiei-Ravandi},
  {Rahman}, {Sand}, {Shin}, {Scholz}, {Stairs}, {Smith}, {Su}, \&
  {Tendulkar}}]{Mckinven2023ApJ}
{Mckinven}, R., {Gaensler}, B.~M., {Michilli}, D., {et~al.} 2023, \apj, 950,
  12, \dodoi{10.3847/1538-4357/acc65f}

\bibitem[{{Men} \& {Barr}(2024)}]{2024A&A...683A.183M}
{Men}, Y., \& {Barr}, E. 2024, \aap, 683, A183,
  \dodoi{10.1051/0004-6361/202348247}

\bibitem[{{Michilli} {et~al.}(2018){Michilli}, {Seymour}, {Hessels}, {Spitler},
  {Gajjar}, {Archibald}, {Bower}, {Chatterjee}, {Cordes}, {Gourdji}, {Heald},
  {Kaspi}, {Law}, {Sobey}, {Adams}, {Bassa}, {Bogdanov}, {Brinkman},
  {Demorest}, {Fernandez}, {Hellbourg}, {Lazio}, {Lynch}, {Maddox}, {Marcote},
  {McLaughlin}, {Paragi}, {Ransom}, {Scholz}, {Siemion}, {Tendulkar}, {van
  Rooy}, {Wharton}, \& {Whitlow}}]{2018Natur.553..182M}
{Michilli}, D., {Seymour}, A., {Hessels}, J.~W.~T., {et~al.} 2018, \nat, 553,
  182, \dodoi{10.1038/nature25149}

\bibitem[{{Minter} \& {Spangler}(1996)}]{1996ApJ...458..194M}
{Minter}, A.~H., \& {Spangler}, S.~R. 1996, \apj, 458, 194,
  \dodoi{10.1086/176803}

\bibitem[{{Nimmo} {et~al.}(2025){Nimmo}, {Pleunis}, {Beniamini}, {Kumar},
  {Lanman}, {Li}, {Main}, {Sammons}, {Andrew}, {Bhardwaj}, {Chatterjee},
  {Curtin}, {Fonseca}, {Gaensler}, {Joseph}, {Kader}, {Kaspi}, {Lazda},
  {Leung}, {Masui}, {Mckinven}, {Michilli}, {Pandhi}, {Pearlman},
  {Rafiei-Ravandi}, {Sand}, {Shin}, {Smith}, \& {Stairs}}]{2025Natur.637...48N}
{Nimmo}, K., {Pleunis}, Z., {Beniamini}, P., {et~al.} 2025, \nat, 637, 48,
  \dodoi{10.1038/s41586-024-08297-w}

\bibitem[{{Niu} {et~al.}(2022){Niu}, {Zhu}, {Zhang}, {Yuan}, {Zhou}, {Zhang},
  {Jiang}, {Han}, {Li}, {Lee}, {Wang}, {Feng}, {Li}, {Luo}, {Wang}, {Dai},
  {Miao}, {Niu}, {Xu}, {Zhang}, {Wang}, {Wang}, \& {Xu}}]{Niu2022RAA}
{Niu}, J.-R., {Zhu}, W.-W., {Zhang}, B., {et~al.} 2022, Research in Astronomy
  and Astrophysics, 22, 124004, \dodoi{10.1088/1674-4527/ac995d}

\bibitem[{{Niu} {et~al.}(2024){Niu}, {Wang}, {Jiang}, {Qu}, {Zhou}, {Zhu},
  {Lee}, {Han}, {Zhang}, {Li}, {Cao}, {Fang}, {Feng}, {Fu}, {Jiang}, {Jing},
  {Li}, {Li}, {Luo}, {Meng}, {Miao}, {Miao}, {Niu}, {Pan}, {Wang}, {Wang},
  {Wang}, {Wang}, {Wu}, {Wu}, {Xu}, {Xu}, {Xu}, {Xue}, {Yang}, {Yuan}, {Yue},
  {Zhao}, {Zhang}, {Zhang}, {Zhang}, {Zhang}, {Zhang}, \&
  {Zhu}}]{2024ApJ...972L..20N}
{Niu}, J.~R., {Wang}, W.~Y., {Jiang}, J.~C., {et~al.} 2024, \apjl, 972, L20,
  \dodoi{10.3847/2041-8213/ad7023}

\bibitem[{{Petroff} {et~al.}(2019){Petroff}, {Hessels}, \&
  {Lorimer}}]{Petroff19AAR}
{Petroff}, E., {Hessels}, J.~W.~T., \& {Lorimer}, D.~R. 2019, \aapr, 27, 4,
  \dodoi{10.1007/s00159-019-0116-6}

\bibitem[{{Piro} \& {Gaensler}(2018)}]{Piro2018}
{Piro}, A.~L., \& {Gaensler}, B.~M. 2018, \apj, 861, 150,
  \dodoi{10.3847/1538-4357/aac9bc}

\bibitem[{{Rajwade} \& {van den Eijnden}(2023)}]{2023A&A...673A.136R}
{Rajwade}, K.~M., \& {van den Eijnden}, J. 2023, \aap, 673, A136,
  \dodoi{10.1051/0004-6361/202245468}

\bibitem[{{Scargle}(1982)}]{Scargle1982}
{Scargle}, J.~D. 1982, \apj, 263, 835, \dodoi{10.1086/160554}

\bibitem[{{Torrence} \& {Compo}(1998)}]{Torrence1998}
{Torrence}, C., \& {Compo}, G.~P. 1998, Bulletin of the American Meteorological
  Society, 79, 61, \dodoi{10.1175/1520-0477(1998)079<0061:APGTWA>2.0.CO;2}

\bibitem[{{Wang} {et~al.}(2011){Wang}, {Han}, \& {Lai}}]{2011MNRAS.417.1183W}
{Wang}, C., {Han}, J.~L., \& {Lai}, D. 2011, \mnras, 417, 1183,
  \dodoi{10.1111/j.1365-2966.2011.19333.x}

\bibitem[{{Wang} {et~al.}(2022){Wang}, {Zhang}, {Dai}, \&
  {Cheng}}]{Wang2022NatCo}
{Wang}, F.~Y., {Zhang}, G.~Q., {Dai}, Z.~G., \& {Cheng}, K.~S. 2022, Nature
  Communications, 13, 4382, \dodoi{10.1038/s41467-022-31923-y}

\bibitem[{{Wang} {et~al.}(2023){Wang}, {Wang}, {Li}, {Yao}, {Manchester},
  {Hobbs}, {Wang}, {Dai}, {Xu}, {Luo}, {Feng}, {Wang}, {Li}, {Yu}, {Du}, {Niu},
  {Zhang}, \& {Zhang}}]{2023ApJ...955...36W}
{Wang}, S.~Q., {Wang}, J.~B., {Li}, D.~Z., {et~al.} 2023, \apj, 955, 36,
  \dodoi{10.3847/1538-4357/acea81}

\bibitem[{{Xu} {et~al.}(2022){Xu}, {Niu}, {Chen}, {Lee}, {Zhu}, {Dong},
  {Zhang}, {Jiang}, {Wang}, {Xu}, {Zhang}, {Fu}, {Filippenko}, {Peng}, {Zhou},
  {Zhang}, {Wang}, {Feng}, {Li}, {Brink}, {Li}, {Lu}, {Yang}, {Caballero},
  {Cai}, {Chen}, {Dai}, {Djorgovski}, {Esamdin}, {Gan}, {Guhathakurta}, {Han},
  {Hao}, {Huang}, {Jiang}, {Li}, {Li}, {Li}, {Li}, {Li}, {Liu}, {Luo}, {Men},
  {Niu}, {Peng}, {Qian}, {Song}, {Stern}, {Stockton}, {Sun}, {Wang}, {Wang},
  {Wang}, {Wang}, {Wu}, {Xiao}, {Xiong}, {Xu}, {Xu}, {Yang}, {Yang}, {Yao},
  {Yi}, {Yue}, {Yu}, {Yu}, {Yuan}, {Zhang}, {Zhang}, {Zhang}, {Zhao}, {Zheng},
  {Zhu}, \& {Zou}}]{Xu2022Nature}
{Xu}, H., {Niu}, J.~R., {Chen}, P., {et~al.} 2022, \nat, 609, 685,
  \dodoi{10.1038/s41586-022-05071-8}

\bibitem[{{Yang} {et~al.}(2023){Yang}, {Xu}, \& {Zhang}}]{2023MNRAS.520.2039Y}
{Yang}, Y.-P., {Xu}, S., \& {Zhang}, B. 2023, \mnras, 520, 2039,
  \dodoi{10.1093/mnras/stad168}

\bibitem[{{Zhang}(2018)}]{2018ApJ...854L..21Z}
{Zhang}, B. 2018, \apjl, 854, L21, \dodoi{10.3847/2041-8213/aaadba}

\bibitem[{{Zhang}(2020{\natexlab{a}})}]{Zhang20Nature}
---. 2020{\natexlab{a}}, \nat, 587, 45, \dodoi{10.1038/s41586-020-2828-1}

\bibitem[{{Zhang}(2020{\natexlab{b}})}]{2020ApJ...890L..24Z}
---. 2020{\natexlab{b}}, \apjl, 890, L24, \dodoi{10.3847/2041-8213/ab7244}

\bibitem[{{Zhang} {et~al.}(2022){Zhang}, {Wang}, {Feng}, {Zhang}, {Li}, {Tsai},
  {Niu}, {Luo}, {Yao}, {Zhu}, {Han}, {Lee}, {Zhou}, {Niu}, {Jiang}, {Wang},
  {Zhang}, {Xu}, {Wang}, \& {Xu}}]{Zhang2022RAA}
{Zhang}, Y.-K., {Wang}, P., {Feng}, Y., {et~al.} 2022, Research in Astronomy
  and Astrophysics, 22, 124002, \dodoi{10.1088/1674-4527/ac98f7}

\bibitem[{{Zhou} {et~al.}(2022){Zhou}, {Han}, {Zhang}, {Lee}, {Zhu}, {Li},
  {Jing}, {Wang}, {Zhang}, {Jiang}, {Niu}, {Luo}, {Xu}, {Zhang}, {Wang}, {Xu},
  {Wang}, {Yang}, \& {Feng}}]{Zhou2022RAA}
{Zhou}, D.~J., {Han}, J.~L., {Zhang}, B., {et~al.} 2022, Research in Astronomy
  and Astrophysics, 22, 124001, \dodoi{10.1088/1674-4527/ac98f8}

\end{thebibliography}
\bibliographystyle{aasjournal}



\end{document}